\documentclass[a4paper,11pt]{article}

\usepackage{geometry}										
\usepackage[english]{babel}									
\usepackage[utf8]{inputenc}									
\usepackage[T1]{fontenc}									

\usepackage{amssymb,amsfonts,amsmath,amsthm,mathtools}
\usepackage{lmodern}										

\usepackage{booktabs}
\usepackage{hyphenat}
\usepackage{booktabs}
\usepackage{bbm}
\usepackage{multirow}
\usepackage{algorithm}
\usepackage{algorithmic}
\usepackage{bstnotations}
\usepackage{xcolor}
\usepackage{tikz}
\usepackage{indentfirst}									
\usepackage{caption}
\usepackage{subcaption}
\usepackage{graphicx}
\graphicspath{{./Figures}} 

\usepackage{authblk}

\usepackage{hyperref}
\PassOptionsToPackage{unicode}{hyperref}
\hypersetup{
pdftitle = {Surrogate modeling for uncertainty quantification in nonlinear dynamics},
pdfauthor = {Stefano Marelli, Styfen Sch\"ar, Bruno Sudret},
pdfsubject = {},
pdfkeywords = {Uncertainty quantification, Surrogate models, Dynamics, NARX models},
colorlinks = true,            
citecolor = blue,
linkcolor = blue,
urlcolor  = blue
}

\usepackage[capitalize]{cleveref}				
\creflabelformat{equation}{#2(#1)#3}				
\crefrangelabelformat{equation}{#3(#1)#4 to #5(#2)#6}		
\labelcrefformat{subequation}{#2(#1)#3}				
\labelcrefrangeformat{subequation}{#3(#1)#4 to #5(#2)#6}	

\usepackage{natbib}           
\newcommand{\covY}{{\Sigma}_{\ve{Y}}}

\newcommand{\rff}{^{\textrm{ref}}}

\title{Surrogate modeling for uncertainty quantification in nonlinear dynamics}

\author[1]{Stefano Marelli\thanks{marelli@ibk.baug.ethz.ch}}
\author[1]{Styfen Sch\"ar \thanks{styfen.schaer@ibk.baug.ethz.ch}}
\author[1]{Bruno Sudret\thanks{sudret@ethz.ch}} \affil[1]{Chair
  of Risk, Safety and Uncertainty Quantification, ETH Z\"{u}rich, Switzerland}

\date{July 1st, 2025}

\begin{document}
\maketitle
\begin{abstract}
  When faced with the task of predicting the behavior of a complex system, engineers have to deal with
  incomplete knowledge on its effective operating conditions: the loads it will be subjected to, unexpected
  environmental and climatic condition, manufacturing uncertainties, and more.  As a consequence,
  uncertainty quantification (UQ) has become a staple tool in modern modeling-based engineering.  From
  identifying and characterizing multiple sources of uncertainty, to quantitatively assessing their effect
  on the accuracy and reliability of the corresponding model predictions, UQ offers a vast arsenal of
  algorithms and methodologies.

  Due to the stochastic nature of uncertainty, however, most of these techniques require a large number of
  evaluations of one or more computational models that represent the system.  These models can be
  computationally demanding, especially for complex engineering systems, to the point that the available
  computational resources often constrain the spectrum of feasible UQ analyses.  It is therefore not
  surprising that surrogate models, \ie powerful functional proxies that rely on a relatively small set of
  training data, have taken center stage in the state-of-the art in UQ.

  This book chapter offers a concise review of the state-of-the-art in surrogate modeling for uncertainty
  quantification, with a particular emphasis on the challenging task of approximating the full
  time-dependent response of dynamical systems. Time-dependent problems are classified into several
  categories based on the intrinsic complexity of the input excitation. Correspondingly, the chapter
  presents methods that combine principal component analysis with polynomial chaos expansions, as well as
  approaches involving time warping and nonlinear autoregressive models with exogenous inputs (NARX
  models). Each method is illustrated through simple application examples that help clarify the various
  approaches.
\end{abstract}
\newpage

\tableofcontents
\newpage


\section{Introduction to uncertainty quantification}
\label{sec:IntroUQ}


Computational models used in engineering sciences aim at mimicking the behaviour of a real system \emph{in
  silico} by numerically solving its governing equations, which are usually partial differential
equations. After some time- and space- discretization of the latter, the problem boils down to solving a
possibly large set of linear or nonlinear equations, leading to a vector of quantities of interest (QoI)
(a.k.a. model response) $\ve{y} \in \Rr^{Q}$. The input parameters that characterize the system, \ie its
geometrical dimensions, material properties, external loading or operating conditions, can be gathered in a
vector $\ve{x} \in \cd_{\Ve{X}} \subset \Rr^M$. With this notation, the computational model $\cm$ that
predicts the QoIs for each vector of input parameters can be seen as a mapping
$\cm: \ve{x} \mapsto \ve{y} = \cm(\ve{x})$. In practice, such mapping is rarely analytical, apart for
academic examples. Usually, computing $\ve{y} $ for a given $\ve{x}_0 $ requires running a computer code
(\eg{} a finite element software in engineering mechanics), a single run of which may take
minutes to hours for a high-fidelity model.

In the context of uncertainty quantification, the input parameters are not perfectly known, and modelled
accordingly by a random vector $\Ve{X}$ of prescribed probability density function (PDF) $f_{\Ve{X} }$,
which is built using expert knowledge, existing data (through statistical inference), or both. Uncertainty
propagation aims at characterizing the statistical properties of the resulting response random vector $\Ve{Y} = \cm(\Ve{X} ) $. Monte Carlo simulation, the most intuitive method that allow one to carry out such an analysis, is based on sampling a (usually large) set of input parameters $\cx = \acc{\ve{x}_1 \enu \ve{x}_{\nmcs}}$
and running the model $\cm$ for each sample. 
The resulting set of model responses can then be post-processed to compute statistical moments, quantiles, probabilities of exceedence or sensitivity indices.

Models of dynamical systems, often central to engineering applications, are a particular case of computational models, where $\ve{Y}$, and often also $\ve{X}$, are time-dependent. 
In this case, the model can be seen as a
mapping from the input vector $\ve{x} \in \Rr^{M}$ and time $t \in \Rr$ to the model response
$\ve{y} \in \Rr^{Q}$:
\begin{equation}
  \label{eq:xxx}
  \cm: (\ve{x}, t) \mapsto \ve{y} (t) = \cm(\ve{x},t),
\end{equation}
with $t\in [0, T]$ for applications with finite duration $T$.
In practice, time is discretized to be computationally tractable, with $\ct = \acc{t_1 \enu t_Q}$, leading to a vector model output
$\ve{y} = \prt{\cm(\ve{x},t_1) \enu \cm(\ve{x}, t_Q)}\tr$.

\section{Surrogate models for time-independent problems}
\label{sec:Surrogates}

%
%
\subsection{Classical families of surrogates}
\label{sec:02-1.1}
The use of brute-force approaches such as Monte Carlo simulation for uncertainty propagation, or genetic
algorithms for optimization, is usually intractable when each run of the model $\cm$ already requires a
significant computational time. To bypass this problem, \emph{surrogate models} have emerged in the last
three decades. Surrogate models are analytical functions selected in a certain family, which approximate
the original model while being extremely fast to evaluate. More precisely, a surrogate is a mapping
$\mhat: \ve{x} \in \Rr^M \mapsto \yyhat \in \Rr^Q$ that satisfies (in some sense to be precised):
\begin{equation}
  \label{eq:001}
  \cm(\ve{x}) \approx \mhat(\ve{x}).
\end{equation}
In Table~\ref{tab:01} the most common families of surrogates are listed together with their functional shape
and parameters, namely polynomial chaos expansions (PCE) \citep{Ghanembook2003,Xiu2002,Xiu2010}, low-rank
tensor approximations \citep{Doostan2013,Chevreuil2015,KonakliRESS2016,KonakliJCP2016}, Gaussian processes
(a.k.a. Kriging) \citep{Santner2003,Rasmussen2006}, support vector machines \citep{Smola2004}, and (deep)
neural networks \citep{Murphy2012,Goodfellow2016}. In this chapter, we will focus on polynomial chaos expansions, which are presented in more detail below.

\newcommand{\vli}{v_l^{(i)}}
\newcommand{\zkli}{z_{k,l}^{(i)}}
\begin{table}[!ht]
  \centering \small 
  \caption{Most common families of surrogate models used for uncertainty quantification}
  \label{tab:01}
  \hspace*{-17mm}
  \begin{tabular}[c]{lcc} 
    %
    \hline  {\bf  Name}& {\bf Shape}& {\bf Parameters} \\
    \hline
    Polynomial chaos expansions & $\widehat\cm(\ve{x}) =
      \dsp{\sum_{\ua\in \ca } a_{\ua} \, \Psi_{\ua}} (\ve{x})$ &
      $\ve{a}_{\ua}$ \\
      Low-rank tensor approximations & $\widehat\cm(\ve{x})= \dsp{\sum_{l=1}^
      R b_l \prt{\prod_{i=1}^M  \left(
          \sum_{k=0}^{p_i} \zkli  \phi_k^{(i)}   (x_i)
        \right)}}$ &  $ \ve{b}, \, \ve{z}$\\
      Kriging (a.k.a Gaussian processes) & $\widehat\cm(\ve{x}) =
      \dsp{\ve{\beta}\tr \cdot \ve{f}(\ve{x}) + \sigma_Z Z(\ve{x}, \omega)}$ &
      $\ve{\beta} \,,\, \sigma_Z^2 \,,\, \ve{\theta} $\\
      Support vector machines & $\widehat\cm(\ve{x}) = \dsp{\sum_{i=1}^M
        a_i \, K(\ve{x}_i , \ve{x}) + b}$ & $ \ve{a} \,,\, b $\\
    (Deep) Neural networks &  $\widehat\cm(\ve{x})= f_n\prt{ \cdots f_2\prt{b_2+ f_1\prt{b_1+ \ve{w}_1 \cdot
                             \ve{x} } \cdot
                             \ve{w}_2}}$ &                                                                                      $\ve{w}, \ve{b}$     \\  
    \hline
    \end{tabular}
\end{table}

\subsection{Training and validation of classical surrogates}
\label{sec:02-1.2}
Each type of surrogate depends on parameters generically denoted by $\ve{\theta} \in \Rr^{n_\theta}$ (see
Table~\ref{tab:01}, 3rd column), which are explicitly shown in the notation
$\mhat\prt{\ve{x} \, ; \ve{\theta}}$. The latter are fitted from simulated data, \ie an \emph{experimental
  design} (ED) $\xtrain =\acc{\ve{x}^{(1)} \enu \ve{x}^{(\ned)}}$ and the corresponding model evaluations
$\ytrain =\acc{y^{(1)} \enu y^{(\ned)}}$ with $y^{(i)} = \cm(\ve{x}^{(i)}), \, i=1 \enu \ned$.\footnote{We
  assume here that the output QoI is scalar for simplicity, \ie $Q=1$.} More specifically, a \emph{loss
  function} such as the mean-square error is introduced to quantify the discrepancy between the model and
its surrogate over the DOE, and minimized over the parameters $\ve{\theta}$:
\begin{equation}
  \label{eq:002}
  \begin{split}
  \widehat{\ve{\theta}} =& \mathop{\arg\min}_{\ve{\theta}} \cl\prt{\ytrain, \mhat\prt{\xtrain}}\\ =& 
\mathop{\arg\min}_{\ve{\theta}} %
\frac{1}{\ned}\sum_{i=1}^{\ned}\prt{y^{(i)} - \mhat\prt{\ve{x}^{(i)}\, ;
    \ve{\theta}}}^2.  
  \end{split}
\end{equation}
The accuracy of the surrogate is usually evaluated using a \emph{test set} of $\ntest$ points sampled in the
input space, and their associated model responses.


\subsection{Sparse polynomial chaos expansions}
\label{sec:02-2}
Following the pioneering work of \citet{Ghanembook1991} and \citet{Xiu2002}, polynomial chaos expansions have become one
of the most popular surrogate modeling techniques for uncertainty quantification. Assuming that the input
random vector $\Ve{X}$ has $M$ independent components $\prt{X_1 \enu X_M}$, and that $Y = \cm(\Ve{X}) $ has
a finite variance, the following expansion holds:
\begin{equation}
  \label{eq:003}
  Y = \sum_{\ua \in \Rr^M} c_{\ua} \psi_{\ua}(\Ve{X}),
\end{equation}
where $\ua$ are multi-indices and $\acc{\psi_{\ua}, \, \ua \in \Rr^M}$ are multivariate polynomials in the
input variables defined as
\begin{equation}
  \label{eq:004}
  \psi_{\ua}(\Ve{X}) \eqdef \prod_{i=1}^M \psi_{\alpha_i}^{(i)} (x_i).
\end{equation}
Denoting by $f_{X_i}$ the marginal PDF of the $i$-th input random variable over its support $\cd_{X_i}$, the
univariate polynomials $\acc {\psi_{k}^{(i)}, \, k\in \Nn}$ are orthogonal \textit{w.r.t.} the probability measure
induced by $X_i$ and satisfy:
\begin{equation}
  \label{eq:005}
  \int_{\cd_{X_i}} \psi_{j}^{(i)}(x) \psi_{k}^{(i)}(x) f_{X_i} (x)\ \di x =\delta_{jk},
\end{equation}
where $\delta_{jk}$ is the Kronecker symbol equal to 1 if $j=k$ and 0 otherwise. Due to the tensor product
construction in Eq.~(\ref{eq:004}), the multivariate polynomials inherit from the orthogonality property:
\begin{equation}
  \label{eq:006}
  \Espe{\Ve{X} }{\psi_{\ua}(\Ve{X}) \psi_{\ub}(\Ve{X})} = \int_{\cd_{\Ve{X}}} \psi_{\ua}(\Ve{x})
  \psi_{\ub}(\Ve{x}) f_{\Ve{X}} (\ve{x}) \, \di \ve{x} = \delta_{\ua \ub},
\end{equation}
with $\delta_{\ua \ub}$ = 1 if $\ua = \ub$ and 0 otherwise.

After selecting a truncation scheme $\ca \subset \Rr^M$, \eg, all polynomials with \emph{total degree}
$||\ua||_1 \le p$, the coefficients $\hat{\ve{c}} \eqdef \prt{c_{\ua}, \, \ua \in \ca}$ can be computed by
ordinary least squares. Denoting by $\ve{\Psi}$ the information matrix defined by:
\begin{equation}
  \label{eq:007}
  \ve{\Psi}_{ij} = \Psi_{\ua_j}(\ve{x}^{(i)}), \quad i=1\enu \ned, \quad j=1\enu \card(\ca),
\end{equation}
where $\ua_j$ is the $j$-th multi-index of $\ca$ following the lexicographic ordering, the PCE coefficients
are obtained from Eq.~(\ref{eq:002}) as:\begin{equation}
  \label{eq:008}
   \hat{\ve{c}} = \mathop{\arg\min}_{c_{\ua} \in \Rr^{\card(\ca)}} \frac{1}{\ned}\sum_{i=1}^{\ned}\prt{y^{(i)} -
    \sum_{\ua \in \ca} c_{\ua}    \psi_{\ua}(\Ve{x}^{(i)})}^2.  
\end{equation}
In practice, the relevant truncation scheme is problem-specific and not known in advance. Modern solvers such
as least-angle regression \citep{Efron_2004,Blatman_2011} extract a sparse expansion from a \emph{candidate}
truncated basis, before the (nonzero) coefficients are computed by ordinary least-squares. For recent
developments on sparse PCEs, the reader is referred to \citet{Luethen2021,LuethenIJUQ2022}.

\section{Surrogate modeling for ``simple'' dynamical systems}
\subsection{Classes of dynamical systems}
\label{sec:surrogates:dynamical}


Extending any of the surrogate modeling techniques in Table~\ref{tab:01} to handle dynamical systems is
usually not straightforward. As a general rule, the complexity of the computational model
(Eq.~\eqref{eq:001}) tends to increase rapidly with time \citep{wan_2005_core, le_2010_core,Choi2014,chu_2016_core}.

We consider here two main classes of problems, distinguished by their input characteristics, that are
treated with entirely different surrogate modeling strategies: those with \emph{fundamentally simple}
inputs, and those with {\it fundamentally complex} inputs. This distinction, first introduced in
\citet{meles2025_core} and \citet{SchaerPhD2025}, refers to the intrinsic dimensionality of the input excitation:
\begin{itemize}
\item \emph{fundamentally simple inputs} can be accurately represented by a finite and fixed number of scalar parameters, typically $O(10)$. 
These can either be time-independent parameter vectors typical of classical surrogate models (see Section~\ref{sec:Surrogates}), or simple time-series with limited frequency content, such as monochromatic sinusoidal waves, or superpositions thereof, which can be easily parameterized with  a small number of frequency amplitudes and phases.
\item \emph{fundamentally complex inputs} do not generally admit compact and/or fixed representations, and generally present a much richer frequency/information content. 
Common examples of this type of inputs are found in stochastic earthquake models \citep{rezaeian2010_core}, or turbulent wind fields \citep{turbsim_2009_mnarx_paper}.
\end{itemize}
The first class of inputs can be generally handled by extending classical  surrogate modeling strategies to vector outputs, in some cases with additional  preprocessing steps to homogenize the input. These approaches are presented in detail in Sections~\ref{sec:02-3} and \ref{sec:TimeWarping}.

The second class of inputs requires instead an entirely different surrogate modeling paradigm, namely auto-regressive modeling, which will be introduced in Section~\ref{sec:UQ:NARX}.

\subsection{Problems with fundamentally simple inputs}
\label{sec:02-3}
The derivation of polynomial chaos expansions of a scalar output quantity of interest
(Section~\ref{sec:02-2}) can be easily extended to a vector $\ve{y}$ with $Q$ components, where the
procedure described in Section~\ref{sec:02-1.2} is simply applied component by component. For dynamical
systems, a naive approach consists in considering each (discretized) output trajectory as a vector (where
each component is attached to a particular time instant of the simulation) and use PCE for each component,
\ie at each time instant. When considering a pre-defined fix basis of polynomials, the solution of the OLS in
Eq.~(\ref{eq:008}) only requires a single matrix inversion, and then a matrix-vector product for each output
component. 
When considering \emph{sparse solvers} though, the optimal sparse basis may be different for each output
$\acc{y_q, q=1 \enu Q}$, leading to a high computational cost when $Q = \co(10^{3-4})$.

Because the computed trajectories for problems with fundamentally simple inputs are similar, a
pre-processing step can be used. \emph{Principal component analysis} \citep{Jolliffe2002} is applied to the
$\ned$ trajectories, and only a small number of components $m \ll min(\ned, Q)$ allows us to represent the
data accurately.

Given the experimental $(\ned \times {Q})$-matrix
$ \ytrain\equiv \prt{\ve{y}^{(1)}; \dots; \ve{y}^{(\ned)}}$, the principal component analysis is carried
out as follows:
\begin{itemize}\itemsep0em
\item The \emph{mean output} $\bar{\ve{y}}$ and \emph{empirical covariance matrix} $\widehat{\covY}$ are
  first computed:
  \begin{equation}
    \label{eq:099}
    \begin{split}
      \bar{\ve{y}} &= \frac{1}{\ned}\sum_{i=1}^{\ned} \ve{y}^{(i)}, \\
      \prt{\widehat{\covY}}_{\!\!ij} &=
      \frac{1}{\ned-1} \sum_{p=1}^{\ned} \prt{y_i^{(p)} - \bar{y}_i} \prt{y_j^{(p)} - \bar{y}_j}.
    \end{split}
  \end{equation}
\item The eigenvalue decomposition of matrix $\widehat{\covY}$ is carried out. The eigenvectors
  (``eigenmodes'') are denoted by $\acc{\phi_j, \, j=1 \enu Q}$. By construction they are normalized. Only
  the $m$ largest eigenvalues and related eigenmodes are kept.
\item For each original $Q$-dimensional vector $\ve{y}^{(i)}$, the \emph{score coefficients}
  $\acc{ z_j^{(i)}, \, j=1\enu m}$ are computed by \emph{projection} onto the eigenmodes
  $\acc{\phi_j, \, j=1\enu m}$:
  \begin{equation}
    \label{eq:098}
    z_j^{(i)} = \phi_j\tr \prt{\ve{y}^{(i)} - \bar{\ve{y}}}.
  \end{equation}
  Note that these score coefficients have a zero mean value by construction. They correspond to the projection coefficients of the original response vector onto the new basis defined by the eigenvectors of $\covY$.
\end{itemize}
From this projection, we obtain a new dataset comprising $\ned$ realizations of score vectors of
length $m$. Independent sparse polynomial chaos expansions of each variable $\acc{Z_j, \, j=1\enu m}$ are then
computed:
\begin{equation}
  \label{eq:097}
  Z_j \approx \sum_{\ua \in  \ca_j} \zeta_{\ua,j} \, \Psi_{\ua}(\Ve{X}),
\end{equation} 
where $\zeta_{\ua,j}$ is the PCE coefficient of polynomial $\Psi_{\ua}$ of the $j$-th score variable $Z_j$.
Finally, these PCEs can be recombined to provide a prediction for the whole curve, for a new input vector
$\ve{x}^*$:
\begin{equation}
  \label{eq:096}
  \widehat {\cm}(\ve{x}^*)= \bar{\ve{y}} + \sum_{j=1}^m \sum_{\ua \in
    \ca_j} \zeta_{\ua,j} \, \Psi_{\ua}(\ve{x}^*) \, \phi_j.
\end{equation}
The number $m$ of components retained is usually chosen so as to explain the $(1-\epsilon)$ fraction of the
empirical variance of the experimental matrix, with $\epsilon = 1-5\%$. This is obtained by taking the $m$ 
largest eigenvalues so that their sum is greater than $(1-\epsilon)\cdot \text{trace}\prt{\widehat{\covY}}$

Such a procedure was first introduced by \citet{BlatmanIcossar2013} together with sparse PCEs. It is
generally known as \emph{model order reduction}, \emph{proper orthogonal decomposition} or simply
\emph{singular value decomposition}. Applications to time-dependent problems with fundamentally simple
inputs can be found in \citet{NagelEAWAG2020} and \citet{WagnerEngStruc2020}, among others.  Recent works also use PCA
in combination with other surrogates to handle problems with very high-dimensional, correlated responses,
see, {\it e.g.} \citet{Aversano2019} and \citet{Guo2023PCA}.

\subsection{Time warping}
\label{sec:TimeWarping}
%
\subsubsection{Introduction}
\begin{figure}[!ht]
  \centering
  \begin{subfigure}[t]{0.48\textwidth}
    \centering
    \includegraphics[width=\textwidth]{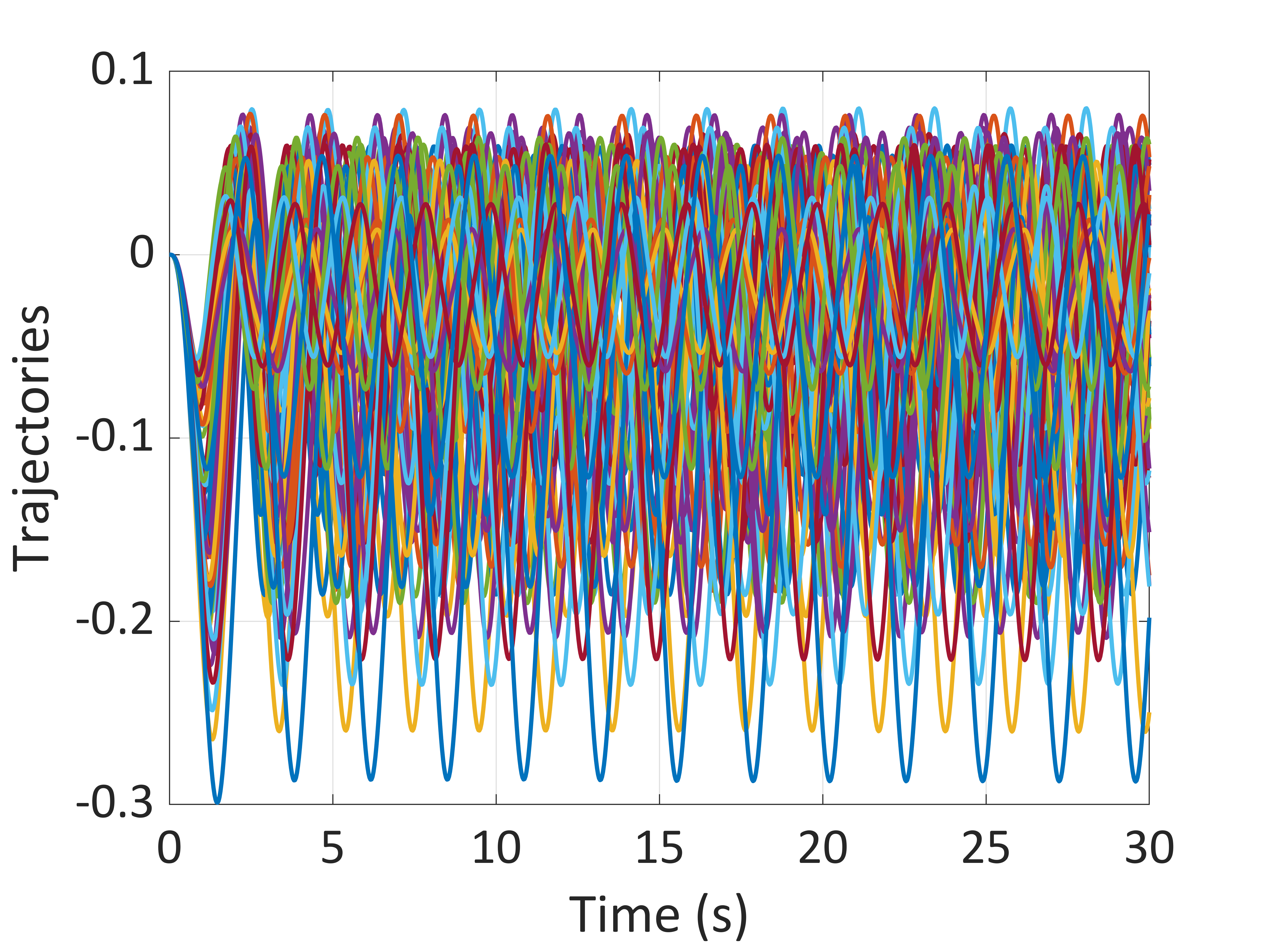}
    \caption{A hundred trajectories in the original (common) time scale $t$}
    \label{fig:001a}
  \end{subfigure}
   \begin{subfigure}[t]{0.48\textwidth}
    \centering
    \includegraphics[width=\textwidth]{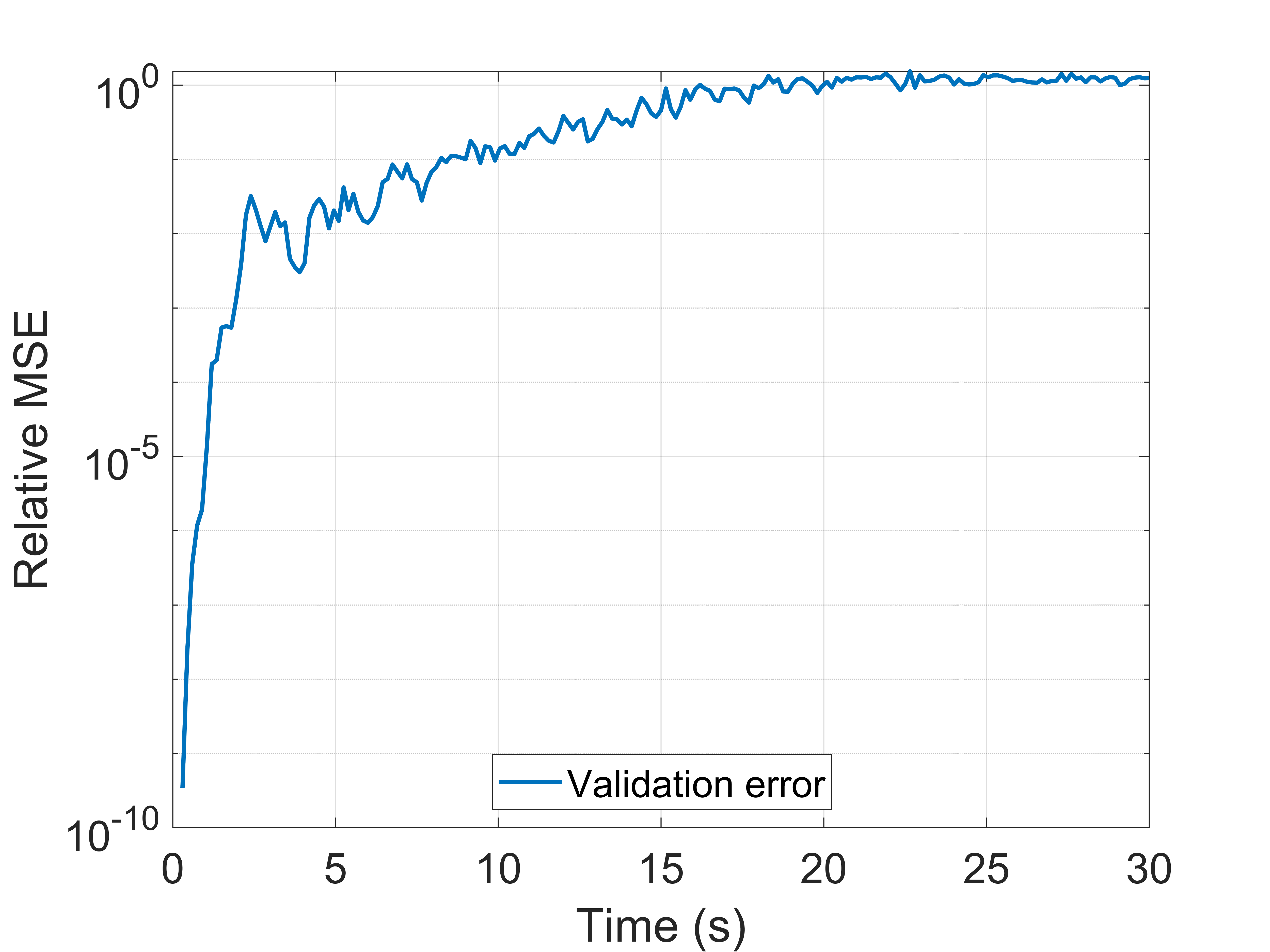}
    \caption{Evolution of the validation error using time-frozen polynomial chaos expansion}
    \label{fig:001b}
  \end{subfigure}
  \caption{Bouc-Wen oscillator with random input parameters under sinusoidal forcing (after
    \citet{MaiSIAMUQ2017}).}
      \label{fig:001}
\end{figure}

The combination of principal component analysis and sparse polynomial chaos expansion presented in the
previous section gives accurate results when the set of training trajectories show some simple common
structure. However, when considering as an example (damped) oscillating systems whose characteristics are
uncertain, the output trajectories may typically look like the ones in Figure~\ref{fig:001a}
(\citet{MaiSIAMUQ2017}): each curve oscillates at its own frequency and phase, so that when considering a particular \emph{frozen} time instant $t^*$, the outputs
$\acc{y^{(i)}(t^*), \, i =1\enu \ned}$ show a complex distribution, since some realizations correspond to a
peak of the trajectory, some to a valley, and others to anything in between \citep{chu_2016_core}. 
When applying sparse PCE at each time instant (\textit{i.e.}, considering each trajectory as a $Q$-dimensional vector and using a PCE for each
component as outlined in Section~\ref{sec:02-3}), which we call \emph{time-frozen} PCE, the point-in-time validation
error rapidly blows up, as shown in Figure~\ref{fig:001b}. Note that the pre-processing step using principal
component analysis, which is a mere \emph{linear transform}, does not solve the problem.

To capture the increasing complexity of the input-output mapping at large time horizons, diverse intrusive
and nonintrusive techniques have been proposed
\citep{Sapsis2009,Gerritsma2010,le_2010_core,Choi2014,Luchtenburg2014}. In this section, we present a fully
nonintrusive technique based on \emph{stochastic time warping}, as developed in \citet{MaiSIAMUQ2017},
following an original idea by \citet{le_2010_core} in an intrusive context.

\subsubsection{Stochastic time warping}
Stochastic time warping can be seen as a preprocessing of the training trajectories in which each curve is
\emph{warped} so as to become ``similar'' to a reference curve $y\rff$, which is \eg{}, the curve obtained
from the mean-value input vector. Practically, in case of oscillatory systems, each trajectory
$y^{(i)}(t) = \cm(t, \ve{x}^{(i)})$ will be displayed along a new time axis $\tau(t,\ve{x}^{(i)})$, where
the mapping $t \mapsto \tau = TW(t,\ve{x}^{(i)})$ is invertible for each realization. This time warping
function $TW$ is parametrized with a few coefficients, \eg{}, a linear combination of given functions $f_j(t)$:
\begin{equation}
  \label{eq:014}
 TW(t, \ve{x}) = \sum_{j=1}^{n_{\beta}} \beta_j(\ve{x}) f_j(t).
\end{equation}
The \emph{warped curve}, for a set of parameters $\ve{\beta}$, is defined as:
\begin{equation}
  \label{eq:015}
  y^{(i)}(\tau; \, \ve{\beta}) = \cm(TW^{-1}(\tau, \ve{x}^{(i)}; \, \ve{\beta}), \ve{x}^{(i)}).
\end{equation}
For each curve $i$, the optimal set of parameters $\ve{\beta}_i^*$ is computed so as to minimize a distance
between the reference curve and the current warped realization:
\begin{equation}
  \label{eq:016}
 {\ve{\beta}}^{(i)} = \mathop{\arg\min}_{\ve{\beta} \in \Rr^{|\ve{\beta}|}} d\prt{y^{(i)}(\tau; \, \ve{\beta}) , \, y\rff}.
\end{equation}
In the original publication of \citet{MaiSIAMUQ2017}, the time warping function is affine:
$ TW(t, \ve{x}) =kt + \phi$, that is $\ve{\beta} = \prt{k, \, \phi}$ or even linear
$ TW(t, \ve{x}) =kt, \; \beta \equiv k$, but more general formats could be used. The norm selected to measure the distance between two (warped)
curves $y_1$ and $y_2$ was the cross-correlation over the time interval of interest $[0,T]$:
\begin{equation}
  \label{eq:017}
  d(y_1(\cdot) , y_2(\cdot)) =
  \frac{\left| \int_0^T y_1(s) \, y_2(s) \, \mathrm{d}s \right|} 
  {\sqrt{\int_0^T y_1^2(s) \, \mathrm{d}s} \,\sqrt{\int_0^T y_2^2(s) \, \mathrm{d}s}}.
\end{equation} 
Note that time warping usually rescales each trajectory on a different time interval, which requires
reinterpolating them on a common time grid over a joint interval $[0,T]$, which requires explicit handling to ensure consistency (see \citet{MaiSIAMUQ2017} for technical details).

The effect of applying time-warping to the Bouc-Wen non-linear oscillator example in Figure~\ref{fig:001a} is shown in Figure~\ref{fig:002a}. After warping, the training curves show remarkable similarity: sparse PCE could then be used at each time
instant, as described in Section~\ref{sec:02-3}. For efficiency, principal component analysis can also be used on the warped curves prior to applying sparse PCE to their score coefficients.

The goal of constructing surrogates is to be able to rapidly emulate new trajectories for new input vectors
$\ve{x}^*$. From the above construction, the prediction occurs \emph{in the warped timescale}, meaning that the
inverse of Eq.~(\ref{eq:014}) should be used to compute the predicted trajectory \emph{in the physical
  time}. To this aim, the warping coefficients $\ve{\beta}(\ve{x} ^*)$ are necessary. This is achieved by
building sparse polynomial chaos expansions of each $\beta_j $ using the experimental design of warping
parameters $\cy_{\ve{\beta}}$ obtained from Eq.~(\ref{eq:016}) from each trajectory:
\begin{equation}
  \label{eq:018}
  \beta_j(\Ve{X})  \approx \sum_{\ua \in  \ca_j} b_{\ua,j} \, \Psi_{\ua}(\Ve{X}), \; j=1 \enu n_\beta.
\end{equation}
As a summary, the time-warping-PCE surrogate modeling technique allows us to emulate the trajectory
$y(\ve{x} ^*)$ as follows:
\begin{itemize} 
\item Compute the coefficients of the time warping function $TW$ for $\ve{x} ^*$ as
  $ \beta_j(\ve{x}^*) \approx \sum_{\ua \in \ca_j} b_{\ua,j} \, \Psi_{\ua}(\ve{x}^*)$
\item Compute the \emph{warped} trajectory $y^*(\tau; \, \ve{\beta}^*)$ using PCA and sparse PCE, as in
  Eq.~(\ref{eq:096})
\item Use the inverse of the time warping function to get the trajectory $y^*\prt{TW^{-1}(\tau; \,
    \ve{\beta}^*)}$ in the physical time scale.
\end{itemize}

\subsubsection{Application: nonlinear Bouc-Wen oscillator}
To illustrate the whole approach, let us consider the single degree of freedom (SDOF) Bouc--Wen oscillator
subject to a stochastic excitation. The equation of motion of the oscillator reads:
\begin{equation}
  \label{eq:019}
\left\{
\begin{aligned}
&\ddot{y}(t) + 2 \zeta \omega \dot{y}(t) + \omega^2 \left( \rho y(t) + (1 - \rho) z(t) \right) = -x(t), \\
&\dot{z}(t) = \gamma \dot{y}(t) - \alpha \left| \dot{y}(t) \right| \left| z(t) \right|^{n-1} z(t) - \beta \dot{y}(t) \left| z(t) \right|^n,
\end{aligned}
\right.
\end{equation}
in which \( \zeta \) is the damping ratio, \( \omega \) is the fundamental frequency, \( \rho =0\) is the
post- to pre-yield stiffness ratio, \( \gamma=1, \, n=1, \alpha,\, \beta \) are parameters governing the
hysteretic loops, and the excitation \( x(t) \) is a sinusoidal function given by
\( x(t) = A \sin(\omega_x t) \). The parameters \(\ve{X} = (\zeta, \omega, \alpha, A, \omega_x) \) are
considered independent random variables, with associated distributions given in
Table~\ref{tab:00x}. A hundred realizations of trajectories have been already shown in Figure~\ref{fig:001a}. 

\begin{table}[!ht]
\centering
\caption{Uncertain parameters of the Bouc--Wen model.}
\label{tab:00x}
\begin{tabular}{llll}
\hline
\textbf{Parameter} & \textbf{Distribution} & \textbf{Mean} & \textbf{Std. dev.}  \\
\hline
\( \zeta \)     & Uniform & 0.02    & 0.002    \\
\( \omega \)    & Uniform & \( 2\pi \)  & \( 0.2\pi \)\\
\( \alpha \)    & Uniform & 50      & 5          \\
\( A \)         & Uniform & 1       & 0.1       \\
\( \omega_x \)  & Uniform & \( \pi \)   & \( 0.1\pi \) \\
\hline
\end{tabular}
\end{table}

\begin{figure}[!ht]
  \centering
     \begin{subfigure}[t]{0.48\textwidth}
       \centering \includegraphics[width=\textwidth]{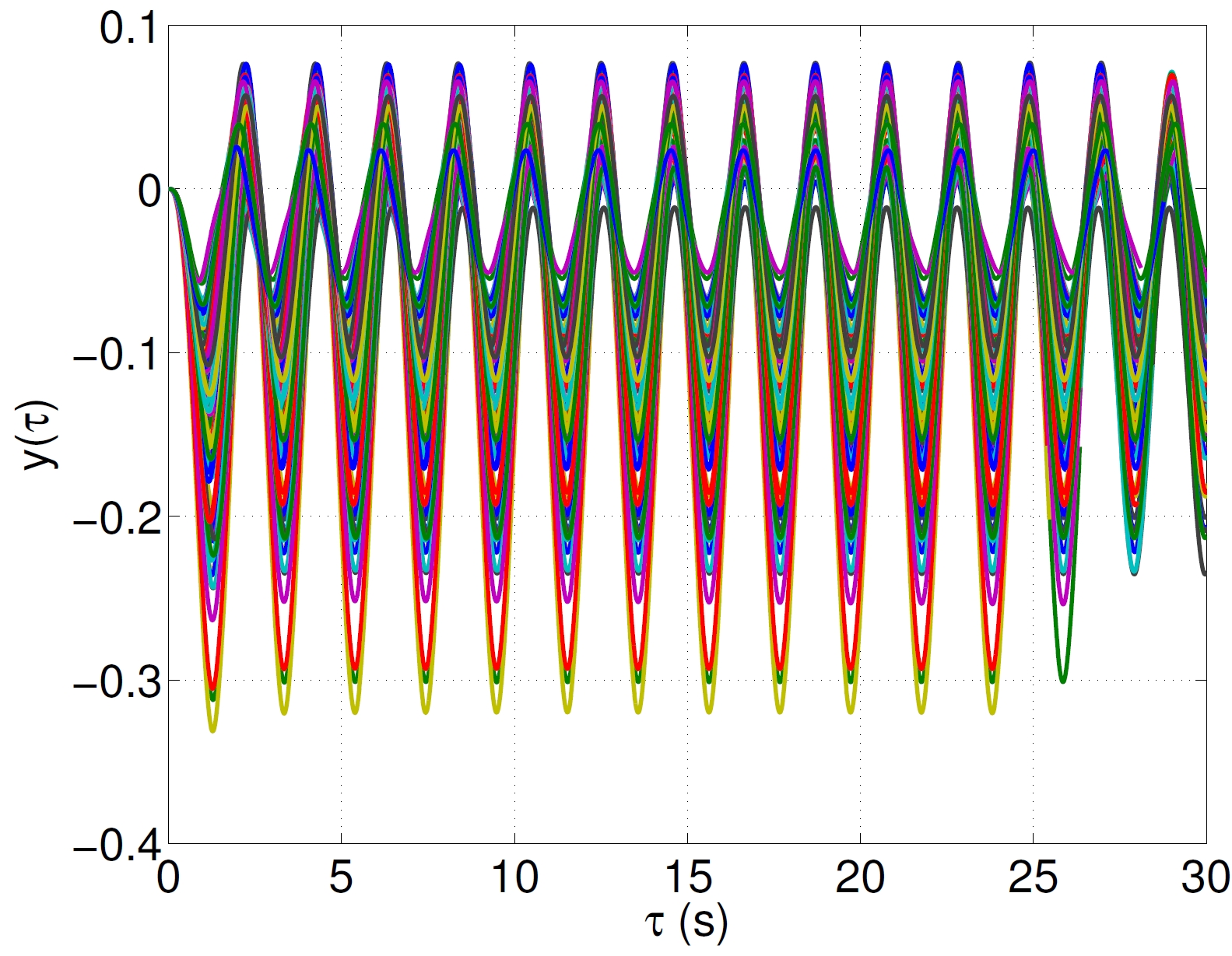}
       \caption{A hundred trajectories in the warped timescale}
    \label{fig:002a}
  \end{subfigure}
     \begin{subfigure}[t]{0.48\textwidth}
    \centering
    \includegraphics[width=\textwidth]{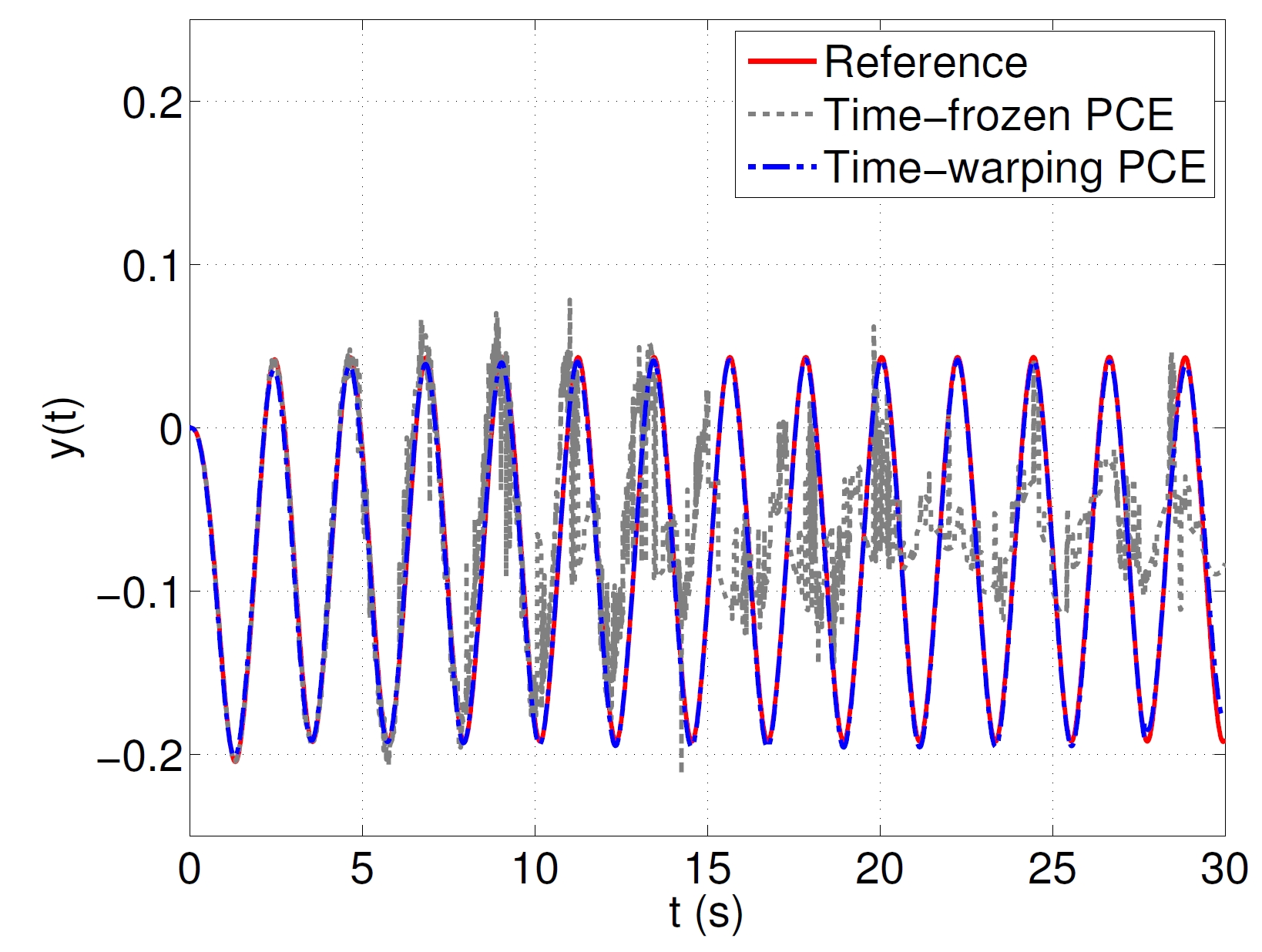}
    \caption{One particular trajectory for ${\ve{x} ^* =\prt{0.019; 5.62; 57.36; 0.94; 2.86}}$ }
    \label{fig:002d}
  \end{subfigure}

     \begin{subfigure}[t]{0.48\textwidth}
       \centering \includegraphics[width=\textwidth]{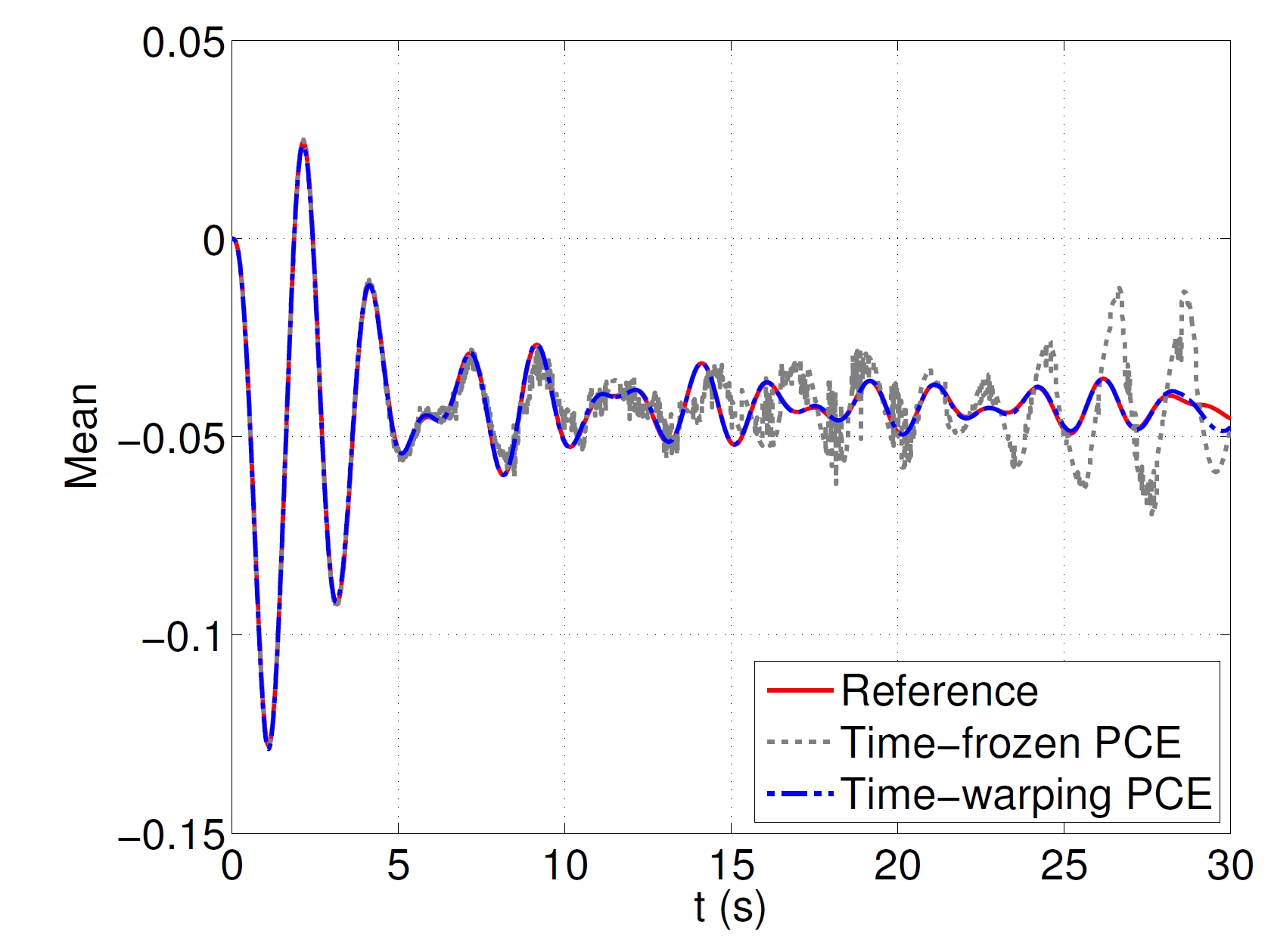}
    \caption{Mean curve}
    \label{fig:002b}
  \end{subfigure}
   \begin{subfigure}[t]{0.48\textwidth}
       \centering \includegraphics[width=\textwidth]{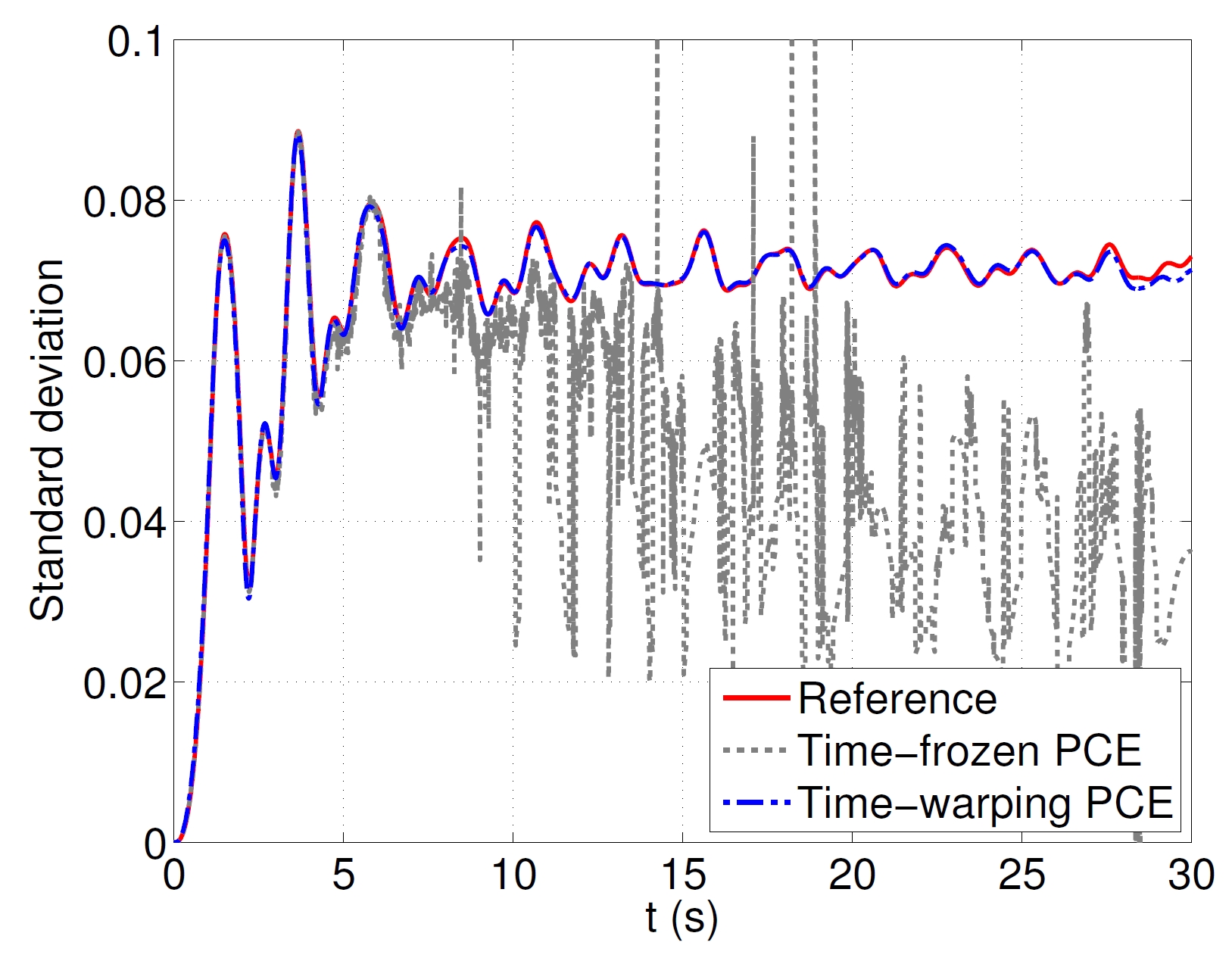}
    \caption{Standard deviation curve}
    \label{fig:002c}
  \end{subfigure}
  \caption{Surrogate model of Bouc-Wen oscillator trajectories with random parameters- and sinusoidal
    forcing (after \citet{MaiSIAMUQ2017}).}
      \label{fig:002}
    \end{figure}

Using the training curves shown in Figure~\ref{fig:001a}, a linear time warping is applied to each
trajectory, and the single coefficient $\beta$ fitted for each of them is used to build a sparse PCE. The
100~curves after time warping are shown in Figure~\ref{fig:002a}. The surrogate obtained by coupling time
warping, principal component analysis and sparse PCE is used to predict 10{,}000 new trajectories, which are
used to estimate the empirical mean and standard deviation curves, shown in blue in
Figure~\ref{fig:002b}-\ref{fig:002c}.

In these figures, the curves predicted using time-frozen PCE are also
plotted in grey, while reference curves obtained from $10^5$~Monte Carlo samples are also plotted in
red. The agreement between reference and time-warping-PCA-PCE is remarkable, while the time-frozen PCE show
poor results after a few seconds. Finally, it is worth mentioning that this surrogate modeling approach
delivers good accuracy not only for statistics, but also for each curve separately: Figure~\ref{fig:002d}
shows a particular trajectory (plotted in red, obtained for the realization
${\ve{x} ^* =\prt{0.0191; 5.6208; 57.3581; 0.9401; 2.8577}}$ and the two surrogate curves obtained by
time-warping-PCA-PCE and time-frozen PCE.

\subsubsection{Conclusions}
In the case of fundamentally simple inputs, surrogate models of dynamical systems can be constructed using
sparse polynomial chaos expansions (PCE) coupled with appropriate pre-processing of the training
trajectories. When the response trajectories are relatively similar in phase and frequency content,
time-frozen PCE may be used effectively. The introduction of principal component analysis (PCA) does not
improve the accuracy in such cases but can significantly reduce the computational cost: instead
of fitting $Q=\co(10^{3-4})$ sparse PCEs, only a handful of $m=\co(10)$ sparse PCE need to be fitted.

Still, for some fundamentally simple inputs parameterized by a small number of random variables, the
resulting trajectories -- though similar in shape -- may differ greatly in phase and frequency. In such
cases, the \emph{stochastic time warping} pre-processing step significantly improves surrogate accuracy. 
This warping has also been successfully applied in the frequency domain, where the curves of
interest are frequency response functions. For such applications, a simple piecewise linear warping function
is used to align the peaks and valleys of the curves with those of a reference curve, see
\citet{Yaghoubi2017} for details.

\section{Auto-regressive models}
\label{sec:UQ:NARX}




\subsection{Introduction}

Approximating the response of complex systems subject to external, fundamentally complex time-dependent loads (see Section~\ref{sec:surrogates:dynamical}) is a well-known
challenge in many fields of model-based engineering.  Examples include damage detection
\citep{mattson_2006,Gao_2016}, control systems \citep{levin_1996,Hu_2024}, maintenance \citep{langeron_2021,
  Samsuri_2023} and design optimization \citep{Yu_2023,deshmukh_2017}, among others.  More recently,
applications in the fields of uncertainty quantification
\citep{spiridonakos_2015,spiridonakos_2015_PCNARX,mai_2016,schar_2024MSSP} and reliability analysis
\citep{garg_2022,Zhou_2023,Zhang2_2024} have started to emerge, due to the large number of model evaluations required for this type of analysis.

All these applications share a common tool: autoregressive models with exogenous inputs (ARX) \citep{billings_2013}.
ARX models, together with their nonlinear variant NARX, are
based on the assumption that the response of a complex system at any give time $t$ is determined by the exogenous loading and the history of the system.  
Following the notation introduced in \citet{schar_2024MSSP}, and considering a discretized time axis
$\ct = \acc {0, \delta t, 2\delta t, \cdots, (Q-1)\delta t}$ this is formalized as:
\begin{equation}\label{eqn:UQ:NARX definition}
    y(t) = \mathcal{M}(\ve{x}(\mathcal{T} \le t), \ve{\beta}),
\end{equation}
where a scalar response $y(t)$ is represented by a deterministic mapping $\cm$ on the $M-$dimensional
exogenous input vector $\ve x (t) \in \mathbb R^M$, and the set of initial system state variables (or boundary conditions) $\ve\beta$. 
For the sake of notational simplicity, we will hereinafter drop the
explicit dependence on $\ve\beta$, unless required by the context.  Additionally, we use the notation
$ \bullet(\mathcal{T} \le t)$ to indicate that the represented quantity depends on the exogenous input only
up to time $t$, but not on future times, to respect causality constraints.

Consistently with the general principles of surrogate modeling, the goal of autoregressive models is to
approximate the response of the original computational model as:
\begin{equation}\label{eq:UQ:dynamic surrogate}
    \widehat{y}(t) = \widehat{\mathcal{M}}(\ve{x}(\mathcal{T} \le t), \ve{\beta}) \approx \mathcal{M}(\ve{x}(\mathcal{T} \le t), \ve{\beta}).
\end{equation}

To achieve the goal of Eq.~\eqref{eq:UQ:dynamic surrogate}, NARX defines the so-called {\it one-step-ahead}
parametric map as:
\begin{equation}\label{eqn:UQ:NARX one step ahead}
    \widehat{y}(t+\delta t) = \widehat{\mathcal{M}}(\ve{x}(\mathcal{T} \le t+\delta t), y(\mathcal{T} < t+\delta t); \ve{c}) + \varepsilon(t),
\end{equation}
where $\varepsilon(t) \sim \mathcal{N}(0, \sigma_{\varepsilon}(t))$ is a zero-mean residual term, often
assumed Gaussian, and $\ve{c}$ is a vector of real parameters that fully characterize the map
$\widehat{\cm}$.  Eq.~\eqref{eqn:UQ:NARX one step ahead} includes all the ingredients at the basis of
autoregressive models with exogenous inputs, as the response at the next time instant $t + \delta t$
depends on the exogenous input excitation until the next time instant $\ve x(\ct\leq t+\delta t)$, but also
on the past responses $\ve y(\ct\leq t)$.

One of the advantages of the formulation in Eq.~\eqref{eqn:UQ:NARX one step ahead} is that there is no
strict requirement on the functional form of the parametric map $\widehat\cm(\bullet; \ve{c})$.  This allows
NARX models to be combined with many other techniques from the statistical regression and machine learning
literature, from classical polynomial regression \citep{billings_2013}, to Gaussian process regression
\citep{murray_1999,kocijan_2012,worden_2018}, neural networks \citep{Siegelmann_1997, li_2021, song_2022},
support vector machines \citep{Gonzalo_2012, rankovic_2014, Zhang_2017}, and many others
\citep{Aguirre_1993, Coca_2001, chen_2008}.

\subsection{Training and prediction with NARX models}\label{sec:UQ:Fitting NARX}

We consider herein nonintrusive NARX models, \textit{i.e.} the surrogate is trained on a finite-size set of full
computational model evaluations, the experimental design $\ve\cx$:
\begin{equation}\label{eqn:UQ:NARX exp design}
    \ve\cx = \left\{ \left( \ve{x}^{(i)}, \ve{y}^{(i)} \right) ,  i=1, \dots, \ned \right\},
\end{equation}
where the $\ve{x}^{(i)} \in \mathbb{R}^{Q \times M}$ represents a collection of input excitations
discretized on the time axis $\ct$, and $\ve{y}^{(i)} = \mathcal{M}(\ve{x}^{(i)}) \in \mathbb{R}^{Q}$ is the
corresponding discretized time response.

The set of coefficients $\ve c$ in Eq~\eqref{eqn:UQ:NARX one step ahead} is determined in a regression
setting.  By combining the set of {\it autoregressive lags} $y(t-(k+1)\delta t)$ and {\it exogenous input
  lags} $x_i(t-k\delta t)$, we can assemble the vector of regressive lags $\ve\varphi(t)$ as:
\begin{equation}\label{eqn:UQ:vatphi}
    \begin{split}
        \ve{\varphi}(t) = \{
        &y(t - \delta t), y(t - 2\delta t), \dots, y(t - n_y\delta t), \\
        &x_1(t), x_1(t - \delta t), \dots, x_1(t - n_{x_1}\delta t), \\ 
        &\dots, \\
        &x_M(t), x_M(t - \delta t), \dots, x_M(t - n_{x_M}\delta t)\},
\end{split}
\end{equation}
where the maximum number of time steps considered for each component $\acc{n_y,n_{x_1},\dots,n_{x_M}}$ are
known as {\it model orders} \citep{billings_2013}.

For a given input/output pair $\acc{\ve x^{(i)}(t), y^{(i)}(t)}$, \eg, a single element of the experimental design
$\ve\cx$ from Eq.~\eqref{eqn:UQ:NARX exp design}, we introduce a so-called {\it design matrix}
$ \ve\Phi \in \mathrm R^{\tilde{Q}\times n}$, with $n=n_y+\sum\limits_{i=1}^{M} n_{x_i}$, which collects all
the vectors of regressive lags $\ve\varphi(t_i)$, as well as the corresponding model responses $y(t_i)$:
\begin{equation}\label{eqn:UQ:Phi_matrix}
    \ve{\Phi} = \begin{pmatrix}
       \ve{\varphi}(t_0) \\
       \ve{\varphi}(t_0+\delta t) \\
       \vdots \\
       \ve{\varphi}(t_0+(Q-1)\delta t)
    \end{pmatrix},
    \quad
    \ve{y} = \begin{pmatrix}
       y(t_0) \\
       y(t_0+\delta t) \\
       \vdots \\
       y(t_0+(Q-1)\delta t)
    \end{pmatrix},
\end{equation}
with $t_0 = \max(n_y, n_{x_1}, \dots, n_{x_M}) \delta t$, and thus $\tilde{Q} = Q-t_0/\delta_t$.

Finally, the design matrices from different input/output pairs from the experimental design in
Eq.~\eqref{eqn:UQ:NARX exp design} can be assembled together into the design matrix $\ve\Phi_\text{ED}$:
\begin{equation}\label{eqn:UQ:large_Phi_matrix}
    \ve{\Phi}_\text{ED} = \begin{pmatrix}
        \ve{\Phi}^{(1)} \\
        \vdots \\
        \ve{\Phi}^{(\ned)}
    \end{pmatrix}, 
    \quad
    \ve{y}_\text{ED} = \begin{pmatrix}
            \ve{y}^{(1)} \\ 
            \vdots \\
            \ve{y}^{(\ned)} 
    \end{pmatrix}.
\end{equation}


It is now possible to estimate the set of model coefficients in Eq.~\eqref{eqn:UQ:NARX one step ahead} by
solving the classical regression problem:
\begin{equation}\label{eqn:UQ:loss_minimization}
  \hat{\ve{c}} = \mathop{\arg\min}_{\ve{c}} \mathcal{L}\left(\ve{y}_\text{ED},
    \widehat{\mathcal{M}}(\ve{\Phi}_\text{ED}; \ve{c})\right), 
\end{equation}
where $\cl(\bullet,\bullet)$ is a suitable loss function, typically a quadratic error term.  Arguably the
most common form of autoregressive models is that of {\it linear-in-the-parameters} models
\citep{billings_2013}, which causes Eq.~\eqref{eqn:UQ:loss_minimization} to take the well-known form of
least squares regression:
\begin{equation}\label{eqn:UQ:least squares}
        \hat{\ve{c}} = \mathop{\arg\min}_{\ve{c}} \| \ve{y}_\text{ED} - \ve{\mathcal{G}}(\ve{\Phi}_\text{ED})\, \ve{c} \|^2,
\end{equation}
where $\ve{\mathcal{G}}: \mathbb R^{|\ve \Phi|} \rightarrow \mathbb R^{|\ve c|}$ is a generally nonlinear
deterministic mapping between lags and regressors, such as multivariate polynomials, or a cosine basis.  In
its basic form, the classical ordinary least square in Eq.~\eqref{eqn:UQ:least squares} admits an explicit
solution:
\begin{equation}\label{eqn:UQ:ordinary least squares}
 \hat{ \ve{c}} = \left( {\ve{\Psi}_\text{ED}}^\top {\ve{\Psi}_\text{ED}} \right)^{-1} {\ve{\Psi}_\text{ED}}^\top
  \ve{y}_\text{ED}, 
\end{equation}
where $\ve{\Psi}_\text{ED} \eqdef \ve{\mathcal{G}}(\ve{\Phi}_\text{ED})$.

Numerous advanced approaches exist for the solution of the optimization problem in Eq.~\eqref{eqn:UQ:least
  squares}, in particular sparsity-inducing regularized regression, also known as compressive sensing, {\it
  e.g.} least absolute shrinkage and selection operator (LASSO, \citet{Tibshirani_1996_LASSO}), or least
angle regression (LARS, \citet{Efron_2004}).

Once the vector of parameters $\hat{\ve{c}}$ is estimated by regression from the available experimental
design, the dynamic response of a model to a new, unseen exogenous input $\ve x^*(t)$ can be predicted
iteratively from the one-step-ahead (OSA) predictor in Eq.~\eqref{eqn:UQ:NARX one step ahead}, a process
known as {\it model forecast}.  More formally:
\begin{equation}\label{eqn:UQ:NARX osa predictor}
    \widehat{y}(t+\delta t) = \ve{\mathcal{G}}(\ve{\varphi}^*(t+\delta t))\, \ve{c},
\end{equation}
where the vector of prediction regressive lags $\ve{\varphi}^*$ is constructed as follows:
\begin{equation}\label{eq:varphi_prediction}
    \begin{split}
            \widehat{\ve{\varphi}}(t+\delta t) = \{
            &\widehat{y}(t), \widehat{y}(t - \delta t), \dots, \widehat{y}(t - (n_y-1)\delta t), \\
            &x_1(t+\delta t), x_1(t), \dots, x_1(t - (n_{x_1}-1)\delta t), \\ 
            &\dots, \\
            &x_M(t+\delta t), x_M(t), \dots, x_M(t - (n_{x_M}-1)\delta t)\},
    \end{split}
\end{equation}
with each autoregressive lag $\widehat{y}(t)$ built directly from Eq.~\eqref{eqn:UQ:NARX osa predictor}.
Because this prediction relies on the availability of a set of autoregressive lags even for the first
timestep, it is common to initialize the first $n_y$ timesteps of $\widehat{y}(t)$ to
$\{\widehat{y}(0), ..., \widehat{y}((n_y-1) \delta t)\} = 0$ or, for validation purposes, to the true model predictions
\citep{schar_2024MSSP}.

\subsection{Limitations of NARX modeling}\label{sec:UQ:NARX limitations}
While NARX has been proven to be accurate in multiple applied settings, it has two major limitations that
can affect its usability in the context of uncertainty quantification:
\begin{itemize}
\item it can only represent a stationary system, {\it i.e.} its coefficients encode the behavior of a
  particular system under variable time-dependent excitations. A new model must be trained if the physical
  properties of the system are changed, even if the same excitations are considered. As an example, a
  classical NARX model cannot approximate the response of an oscillator with variable stiffness or geometry;

\item the input/output map $\widehat\cm (\bullet, \bullet)$ in Eq.~\eqref{eqn:UQ:NARX definition} has in
  general finite complexity/expressivity, which in turn is connected to the size of the parameter vector
  $\ve c$. This makes strongly nonlinear maps difficult to approximate with relatively small experimental
  designs;
    
\item complex computational models may depend on a large number of lags, causing the number of regressors in
  Eq.~\eqref{eqn:UQ:least squares} to become unmanageably large.  This is a common problem in surrogate
  modeling, known as the {\it curse of dimensionality}.
\end{itemize}

In the following sections, we introduce two state-of-the-art methods to solve the first two challenges
in real world applications: polynomial-chaos NARX (PC-NARX,
\citet{spiridonakos_2015,spiridonakos_2015_PCNARX,mai_2016}) to handle model nonstationarity in
Section~\ref{sec:UQ:PC-NARX}, and manifold-NARX (mNARX, \citet{schar_2024MSSP}) to handle both highly nonlinear and high-dimensional maps in Section~\ref{sec:UQ:mNARX}, respectively.
The third challenge, related to the curse of dimensionality, is discussed in depth and tackled through a novel functional reformalization of NARX modeling, namely functional-NARX (F-NARX), in \citet{schar2025FNARX}, which lies outside the scope of this chapter.

\subsection{Polynomial-chaos NARX}\label{sec:UQ:PC-NARX}
If the dynamical response of the system under consideration also depends on a set of $M_{\ve\xi}$ uncertain
{\it structural parameters} $\ve \xi\in \mathbb R^{M_{\ve\xi}}$ with joint probability distribution
$\ve\xi \sim f_{\ve \Xi}$, 
we can rewrite Eq.~\eqref{eqn:UQ:NARX definition} as follows \citep{spiridonakos_2015_PCNARX}:
\begin{equation}\label{eqn:UQ:PC-NARX system}
    y(t) = \mathcal{M}(\ve{x}(\mathcal{T} \le t), \ve \xi, \ve{\beta}).
\end{equation}
Because the components of $\ve\xi$ are not time-dependent, they cannot be used directly in the
one-step-ahead formalism introduced in Eq.~\eqref{eqn:UQ:NARX one step ahead}.  As a consequence,
\citet{spiridonakos_2015_PCNARX,spiridonakos_2015,mai_2016} propose to include the parametric variability directly in the coefficients of the one-step-ahead prediction in Eq.~\eqref{eqn:UQ:NARX one step ahead}, by
modifying it as follows:
\begin{equation}\label{eqn:UQ:PCNARX OSA}
    \widehat{y}(t+\delta t,\ve\xi) = \widehat{\mathcal{M}}(\ve{x}(\mathcal{T} \le t+\delta t), y(\mathcal{T} < t+\delta t), \ve c(\ve\xi)) + \varepsilon(t),
\end{equation}
where the explicit dependence on the initial conditions
$\ve\beta$ has been dropped for notational simplicity.  Intuitively, Eq.~\eqref{eqn:UQ:PCNARX OSA}
represents a family of NARX models parametrized by the structural parameters $\ve\xi$.

With this formalism, \citet{spiridonakos_2015_PCNARX,spiridonakos_2015} and \citet{mai_2016} propose a
two-step approach to surrogate dynamic systems with variable structural parameters:
\begin{itemize}
\item create a surrogate model of the vector of model parameters ${\widehat {\ve c}(\ve\xi)\approx \ve
    c(\ve\xi)}$ based on sparse polynomial chaos expansion \citep{Blatman_2011,mai_2016}:
    \begin{equation}\label{eqn:UQ:PCNARX PCE}
        \widehat c_\kappa = \sum\limits_{j = 1}^{N_{c_\kappa}} a_j^\kappa \psi_j^\kappa (\ve\xi),
    \end{equation}
    where the
    $\psi_j$'s are multivariate polynomials suitably constructed from the joint probability distribution of
    the structural parameters (See Eq.~(\ref{eq:004}));
  \item use the surrogated parameters in the standard NARX one-step-ahead predictor in
    Eq.~\eqref{eqn:UQ:NARX osa predictor}, which now reads:
    \begin{equation}\label{eqn:UQ:PCNARX osa predictor}
        \widehat{y}(t+\delta t) = \ve{\mathcal{G}}(\ve{\varphi}^*(t+\delta t))\, \widehat{\ve{c}}(\ve\xi^*).
    \end{equation}
\end{itemize}
The training of a PC-NARX model from a given experimental design is quite similar to that of a classical
NARX model (see Section~\ref{sec:UQ:Fitting NARX}), with the exception of an additional step to fit the coefficients of the PCE model $\widehat{\ve{c}}(\ve\xi)$. As a first step, the experimental design in Eq.~\eqref{eqn:UQ:NARX exp design} must now include the uncertain structural parameters $\ve\xi$:
\begin{equation}\label{eqn:UQ:PCNARX exp design}
    \ve\cx = \left\{ \left( \ve{x}^{(i)},\ve\xi^{(i)}, \ve{y}^{(i)} \right) ,  i=1, \dots, \ned \right\}.
\end{equation}
The second step is to train a separate autoregressive model {for each element of the experimental design}, without considering the structural parameters $\ve\xi$, by directly solving the least square problem as:
\begin{equation}\label{eqn:UQ:PCNARX ordinary least squares}
    \ve c^{(i)} = \left( {\ve{\Psi}^{(i)}}^\top {\ve{\Psi}^{(i)}} \right)^{-1} {\ve{\Psi}^{(i)}}^\top \ve{y}^{(i)},
\end{equation}
with $\ve{\Psi}^{(i)} \eqdef
\ve{\mathcal{G}}(\ve{\Phi}^{(i)})$.  For details on how to select a common regressor basis
$\mathcal{G}(\bullet)$ for all the traces in the experimental design, an important step to ensure stability
of the algorithm, the reader is referred to \citet{mai_2016}.

Once the regression coefficients for each element of the experimental design is available, all what is
needed to calculate the model predictions in Eq.~\eqref{eqn:UQ:PCNARX osa predictor} are the expansion
coefficients $\ve a^{\kappa} = \acc{a^\kappa_1,\dots,a^\kappa_{N_{c_\kappa}}}$ in Eq.~\eqref{eqn:UQ:PCNARX
  PCE}.  They are calculated separately for each autoregressive coefficient $\widehat c_\kappa (\ve\xi)$
\citep{spiridonakos_2015_PCNARX,mai_2016}.  We start by assembling the following experimental design:
\begin{equation}\label{eqn:UQ:PCNARX ED Matrices}
    \ve{\Xi}_\text{ED} = \begin{pmatrix}
        \ve{\xi}^{(1)} \\
        \vdots \\
        \ve{\xi}^{(\ned)}
    \end{pmatrix}, 
    \quad
    \ve{c}^\kappa_\text{ED} = \begin{pmatrix}
            c_\kappa^{(1)}\\ 
            \vdots \\
            c_\kappa^{(\ned)} 
    \end{pmatrix}.
\end{equation}
Once the experimental design is available, state-of-the-art polynomial chaos expansion can be employed
directly to calculate both the orthogonal polynomials (from the joint distribution of the structural
parameters $f_{\ve\Xi}$) and the coefficients \citep{Blatman_2011,mai_2016, LuethenIJUQ2022} needed to
perform predictions on unseen data, following Eq.~\eqref{eqn:UQ:PCNARX osa predictor}.

\subsection{Manifold-NARX}\label{sec:UQ:mNARX}
Sometimes complex computational models can be strongly nonlinear, even in the absence of structural
parameters, or may require a large number of exogenous inputs. A typical example of both these class of
models is given by aero-servo-elastic simulators, which calculate the dynamic structural or functional
response of a wind turbine, including its control system, to high-dimensional turbulent wind timeseries.

In this class of problems, the direct use of autoregressive models is essentially impossible, because the
topological complexity of the mapping $\widehat\cm(\bullet,\bullet)$ in Eq.~\eqref{eqn:UQ:NARX one step
  ahead} becomes intractable even with the most advanced learners available in the regression and
machine-learning literature.

To address both of these problems, \citet{schar_2024MSSP} recently proposed {\it manifold-NARX} (mNARX), a
variant of NARX that introduces a low-dimensional {\it autoregressive manifold}, on which
$\widehat\cm(\bullet,\bullet)$ becomes tractable.  In more quantitative terms, the idea is that we want to
substitute the original exogenous input $\ve x$ in Eq.~\eqref{eqn:UQ:NARX definition} with a manifold
$\ve\zeta$, such that the new mapping $\tilde\cm\, : \, \ve\zeta(\ct \leq t)\rightarrow y(t)$ is less
nonlinear, resulting in simpler and more accurate autoregressive models.  This manifold is constructed
iteratively by defining a sequence of {auxiliary quantities} $z_i(t)$, either by directly processing the
exogenous input (e.g. through dimensionality reduction techniques), or by introducing additional physically
meaningful intermediate quantities based on prior knowledge of the system \citep{schar_2024MSSP}:
\begin{equation}\label{eqn:UQ:mNARX zeta}
    \begin{split}
        z_{1}(t) &= \mathcal{F}_1(\ve{x}(\mathcal{T} \le t), z_{1}(\mathcal{T} <t)) \\
        {z}_{2}(t) &= \mathcal{F}_2({z}_{1}(\mathcal{T} \le t), \ve{x}(\mathcal{T} \le t), z_{2}(\mathcal{T} <t)) \\
        \vdots \\
        {z}_{i}(t) &= \mathcal{F}_i({z}_{1}(\mathcal{T} \le t), \dots, {z}_{i-1}(\mathcal{T} \le t), \ve{x}(\mathcal{T} \le t), z_{i}(\mathcal{T} <t)),
    \end{split}
\end{equation}
where the $\mathcal{F}_i(\bullet,\bullet)$ represent arbitrarily complex functions.  For example, to
accurately predict the power output and other diagnostic quantities of interest of an operating wind
turbine, \citet{schar_2024MSSP} used as $z_i(t)$ several spatial Fourier modes of exogenous input
(high-dimensional, spatially coherent turbulent wind excitation), as well as additional autoregressive
quantities extracted from the simulation data, such as the turbine's blade pitch (actuated by a controller),
the blade azimuth, and the angular velocity of the rotor.

All of the $z_i(t)$ are then aggregated into the final exogenous input manifold
$\ve\zeta(t) = \acc{z_1(t),\dots,z_{M_{\ve\zeta}}(t)}$, which is substituted to $\ve x$ in
Eq.~\eqref{eqn:UQ:NARX one step ahead}:
\begin{equation}\label{eqn:UQ:mNARX one step ahead}
    \widehat{y}(t+\delta t) = \widehat{\mathcal{M}}(\ve\zeta(\mathcal{T} \le t+\delta t), y(\mathcal{T} < t+\delta t), \ve{c}) + \varepsilon(t).
\end{equation}
From here, the surrogate training procedure is identical to that of classical NARX highlighted in
Section~\ref{sec:UQ:Fitting NARX}.  Predicting the response of an mNARX model on an unseen exogenous input
is also very similar to the case of classical NARX in Eq.~\eqref{eqn:UQ:NARX osa predictor}, with the
additional step of projecting the unseen input onto the autoregressive manifold
$\ve x^*(t) \rightarrow \ve\zeta^*(t)$, through Eq.~\eqref{eqn:UQ:mNARX zeta}.  For an in-depth discussion on the construction, strengths and limitations of mNARX, the reader is referred to \citet{schar_2024MSSP}.

\section{Application to a simple case study: coupled oscillator}
\label{sec:UQ:ARX:case study}
We consider a coupled system with two masses, $m_s$ and $m_u$, connected by a nonlinear spring with
stiffness $k_s$ and a linear damper $c$.  The second mass $m_s$ is also connected to the ground by a linear
spring with stiffness $k_u$.

This system is governed by the following ordinary differential equations:
\begin{equation}\label{eqn:UQ:quarter_car}
  \begin{cases}
& m_u \, \ddot{y}_1(t) = k_s\bigl(y_2(t)-y_1(t)\bigr)^3 + c\bigl(\dot y_2(t) - \dot y_1(t)\bigr) + k_u\bigl(x(t)-y_1(t) \bigl),\\
 & m_s \, \ddot{y}_2(t) = -k_s\bigl(y_2(t)-y_1(t)\bigr)^3 - c\bigl(\dot y_2(t) - \dot y_1(t)\bigr).
  \end{cases}
\end{equation}

The upper mass $m_s$ is significantly smaller than the lower mass $m_u$, and the ratios $\frac{k_s}{m_s}$
and $\frac{k_u}{m_u}$ are of similar magnitude.  Therefore, the displacement of $m_s$ ($y_2$) is highly
dependent on that of $m_u$ ($y_1$).  Conversely, the displacement of $m_u$ is largely unaffected by $m_s$.
The system is subjected to a random excitation on $m_u$ via the lower spring, defined as:
\begin{equation}\label{eq:mass_spring_system_exo_input}
 x(t) = \frac{1}{N_\omega}\sum_{i=1}^{N_\omega} A_i \sin \left( 2\pi B_i t + C_i\right).
\end{equation}

This excitation is the average of $N_\omega$ sinusoidal terms, where $N_\omega$ is a discrete uniform random
variable from 1 to 10. Each term has a random amplitude $A_i \sim \mathcal{U}(-1, 1)$, frequency
$B_i \sim \mathcal{U}(-1, 1)$, and phase $C_i \sim \mathcal{U}(-\pi, \pi)$, resulting in a highly varied
family of signals, including both simple and complex excitations.  Of interest is the displacement of the
top mass $m_s$.

To showcase the different approximation capabilities of polynomial chaos-NARX and Manifold-NARX, we consider
two different system configurations:
\begin{itemize}
\item strong damping with uncertain masses and springs, and
\item weak damping with fixed masses and springs.
\end{itemize}

Each of these cases challenges classical NARX modeling differently: uncertainty in springs and masses
results in variable system responses to the same excitation, breaking classical NARX assumptions on the
fixed properties of the system.  Conversely, weak damping causees the top mass response to be much more
nonlinear, causing classical NARX to require a large number of lags, which can quickly result in an
untreatable number of terms in Eq.~\eqref{eqn:UQ:vatphi}.

\subsection*{High damping and variable system -- PC-NARX}
\label{sec:oscillator:PCNARX}
The first system is characterized by a relatively high damping coefficient, associated with significant
variability in the system parameters.  For reference, the parameter distributions are provided in
Table~\ref{tab:oscillator:PCNARX}.
\begin{table}[!ht]
    \centering
    \caption{Distributions of the parameters of the highly damped oscillator in
      Section~\ref{sec:oscillator:PCNARX}.  All the parameters are normally distributed, with tabulated
      means $\mu_i$ and corresponding coefficients of variation $\sigma_i/\mu_i$.}
    \label{tab:oscillator:PCNARX}
    \begin{tabular}{llll}\hline 
      \textbf{Parameter} & \textbf{Mean} & \textbf{Unit}& \textbf{CoV}$_i = \sigma_i/\mu_i$\\\hline
      $k_u$ & $5{,}000$ & [N/m]  & 0.2\\
      $k_s$ & $1{,}000$ &[N/m$^3$] & 0.2\\
      $m_u$ & $50$ & [kg] & 0.2\\
      $m_s$ & $10$ &[kg] & 0.2\\
      $c$   & $600$ & [N$\cdot$m/s] & 0.2\\
      \hline
    \end{tabular}
\end{table}

Classical NARX modeling is known to perform well in both system identification and surrogate modeling for a
single realization of the system parameters.  Nevertheless, the introduction of variability in the system
parameters is expected to cause classical NARX modeling to be insufficient to capture the complex nonlinear
response of the system in Eq.~\eqref{eqn:UQ:quarter_car}.

In this case, a properly trained PC-NARX (Section~\ref{sec:UQ:PC-NARX}) is expected to dynamically predict
suitable NARX regression coefficients for each realization of the random system parameters, following
Eqs.~\eqref{eqn:UQ:PCNARX PCE} and \eqref{eqn:UQ:PCNARX osa predictor}.

To assess the performance of both NARX and PC-NARX models, we train both models on the same set of
$\ned = 100$ independent realizations of both system parameters and excitation.  For both methods, we
use a polynomial NARX model up to degree $d = 3$, with $n_x = 5$ exogenous input and $n_y = 4$
autoregressive lags, respectively (see Eq.~\eqref{eq:varphi_prediction}).  For classical NARX, the set of
coefficients is calculated via ordinary-least-squares (OLS) over the entire training dataset.  For PC-NARX,
one set of coefficients is calculated for each of the trajectories in the experimental design.  The resulting
coefficients are then surrogated as a function of the input parameters
$\ve{X} = \acc{k_u, k_s, m_u, m_s, c}$ through an adaptive sparse PCE (Eq.~\eqref{eqn:UQ:PCNARX PCE}), with
adaptive truncation up to $d_\text{PCE} = 10$ \citep{LuethenIJUQ2022}.
\begin{figure}[ht]
    \includegraphics[width=\textwidth,clip=true,trim=30 0 75 0]{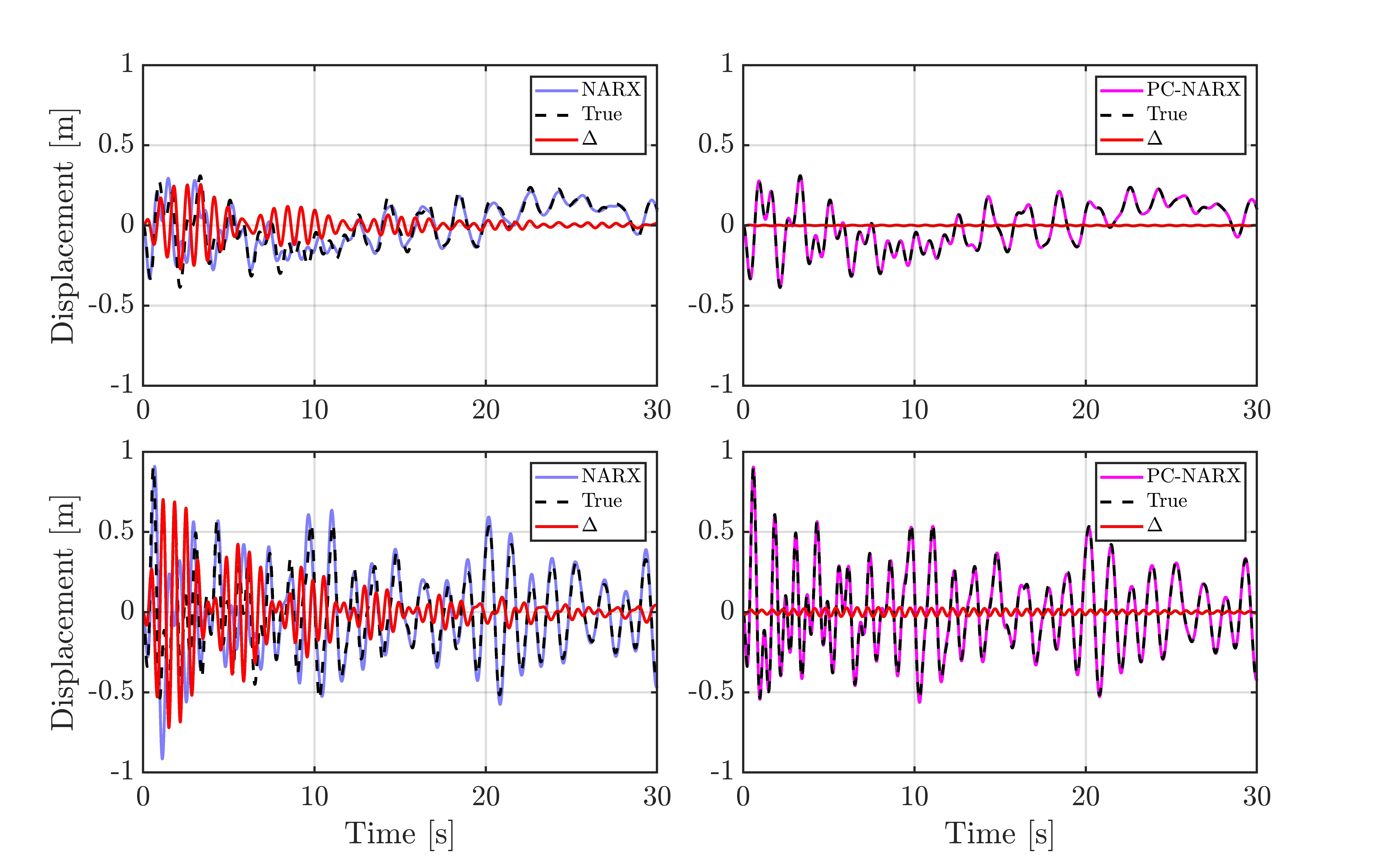}
    \caption{Comparison of classical NARX and PC-NARX surrogates on the strong-damping, uncertain parameters
      case study in Section~\ref{sec:oscillator:PCNARX}. Each row represents an out-of-sample realization of
      both system parameters and exogenous input excitation.  The left column shows the performance of
      classical polynomial NARX, while the right column shows that of PC-NARX. In all plots, the black
      dashed line is the reference validation signal, the blue line represents the NARX approximation, the
      magenta line represents the PC-NARX approximation, and the red line shows the corresponding
      approximation error $\Delta = \widehat y(t) - y(t)$.}
    \label{fig:UQ:PCNARX Traces}
\end{figure}

The comparative performance between the two surrogate modeling approaches is showcased in the two columns of
Figure~\ref{fig:UQ:PCNARX Traces}.  The performance of a classical NARX model trained on the entire dataset
is shown on the left column for two different out-of-sample simulations (rows).  While it generally achieves
discrete approximation performance, the NARX model tends to be inaccurate at the start of the trace, where
the frequency content is generally richer, before the damper has time to dissipate enough energy.  At later
times, the discrepancy between the surrogate and the reference trace decreases significantly.

The performance of a PC-NARX trained on the same dataset is shown on the right column of
Figure~\ref{fig:UQ:PCNARX Traces}.  The surrogate is highly accurate on both traces, even if a decrease in
error over time comparable to classical NARX can still be observed.  Despite the NARX component of PC-NARX
being identical to that of its classical counterpart, the additional fine-tuning of the coefficients brought
by the PCE component of the algorithm allows for a significant improvement in accuracy when the system
parameters are uncertain.

\subsection*{Low damping and fixed system - mNARX}
\label{sec:oscillator:mNARX}

In this second scenario, we consider the same damped oscillator as in the previous section, but with two
significant differences: i) there is no uncertainty on the system parameters, and ii) the damping
coefficient $c$ is strongly reduced.  Their values are reported for reference in
Table~\ref{tab:oscillator:mNARX}.
\begin{table}[!ht]
    \centering
    \caption{Parameters of the weakly damped oscillator in Section~\ref{sec:oscillator:mNARX}.}
    \label{tab:oscillator:mNARX}
    \begin{tabular}{lll} \hline
      Parameter & Value& Unit\\\hline
      $k_u$ & $5{,}000$ & [N/m] \\
      $k_s$ & $1{,}000$ & [N/m$^3$]\\
      $m_u$ & $50$ & [kg]\\
      $m_s$ & $10$  &[kg]\\
      $c$ & $50$ & [N$\cdot$m/s]\\\hline
    \end{tabular}
\end{table}

While removing the variability in the system parameters can in principle simplify the autoregressive
problem, reducing the damping of over an order of magnitude causes the response of the upper mass to become
much more complex to predict.  However, its effect on the lower mass is relatively low, making the system an
ideal candidate for the mNARX algorithm described in Section~\ref{sec:UQ:mNARX}.

To compare the accuracy of classical NARX and mNARX, we proceed as in Section~\ref{sec:oscillator:PCNARX}, by building an experimental design of $\ned = 100$ realizations of the exogenous input excitation, and fitting both models on the same dataset. 
An out-of-sample set of traces is used to test the accuracy of the two methods.

Due to the increased complexity in the displacement of $m_u$, we obtained the best results with a highly
nonlinear NARX with maximum polynomial degree $d = 7$, and using $n_x = 3$ and $n_y = 4$ lags, respectively.

\begin{figure}[!ht]
    \includegraphics[width=\textwidth,clip=true,trim=30 0 75 0]{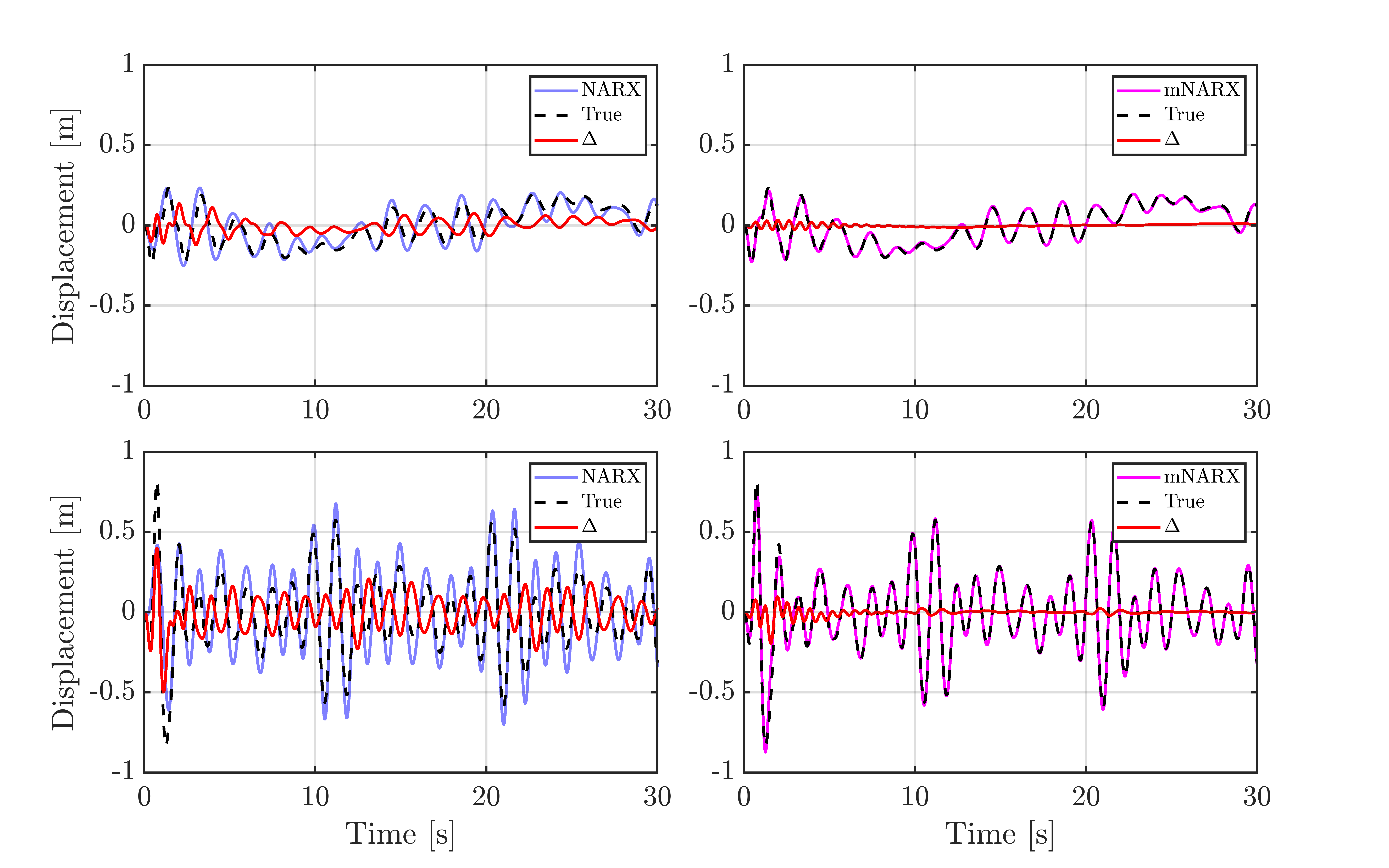}    
    \caption{Comparison of classical NARX and mNARX surrogates on the weak-damping case study in
      Section~\ref{sec:oscillator:PCNARX}. Each row represents an out-of-sample realization of both system
      parameters and exogenous input excitation.  The left column shows the performance of classical
      polynomial NARX, while the right column shows that of mNARX. In all plots, the black dashed line is
      the reference validation signal, the blue line represents the NARX approximation, the magenta line
      represents the mNARX approximation, and the red line shows the corresponding approximation error
      $\Delta = \widehat y(t) - y(t)$.}
    \label{fig:UQ:mNARX Traces}
\end{figure}

The left column of Figure~\ref{fig:UQ:mNARX Traces} shows the performance of classical NARX on two
out-of-sample realizations of the input excitation.  While the overall features of the time series, such as
the dominant frequency and the overall amplitude trends, are correctly approximated, the overall surrogate
accuracy is still rather low.  In contrast to the corresponding plots on the left column of
Figure~\ref{fig:UQ:PCNARX Traces}, no clear trend can be identified in the evolution of the approximation
error over time, as the damping coefficient is too low to effectively dissipate high-frequency energy within
this time scale.

To construct the mNARX model, we take advantage of the of Eq.~\eqref{eqn:UQ:quarter_car}, and use the lower
mass response $y_u(t)$ as an auxiliary quantity in Eq.~\eqref{eqn:UQ:mNARX zeta}.  This allows us to
distribute the autoregressive model for $y_s(t)$ in a sequence of two simpler models:
\begin{equation}
    \begin{split}
        \widehat y_u(t) =&~ \widehat y_u(t,x(\ct \leq t))\\    
        \widehat y_s(t) =&~ \widehat y_s(t,x(\ct \leq t),\widehat y_u(\ct\leq t)).
    \end{split}
    \label{eqn:UQ:Oscillator:Manifold}
\end{equation}
In other words, we enrich the input manifold of the upper mass $m_s$ by including the
autoregressive model of the lower mass $m_u$.  This allows us to create two different autoregressive models,
each with significantly lower complexity than the single NARX, as reported in
Table~\ref{tab:oscillator:mNARX model}.
\begin{table}[!ht]
    \caption{Parameters of the mNARX model for each of the oscillator responses in Eq.~\eqref{eqn:UQ:Oscillator:Manifold}.}
    \label{tab:oscillator:mNARX model}
    \centering
    \begin{tabular}{cccc} \hline
        Response & ~~$d$~~ & ~~$n_x$~~ & ~~$n_y$~~\\\hline
        $y_u$ & $4$ & $2$ & $3$\\
        $y_s$ & $6$ & $2$ & $3$\\\hline
    \end{tabular}
\end{table}
The performance of the resulting surrogate is reported in the right column of Figure~\ref{fig:UQ:mNARX
  Traces}, on the same traces as for the classical NARX counterpart (left column).  Using the much simpler
response of the lower mass $m_u$ highly simplifies the predictions on the response of the second mass,
resulting in an overall dramatic decrease in approximation error, despite the much simpler model.

\section{Conclusions}
Uncertainty quantification has become increasingly widespread in the design and analysis of complex dynamical systems. 
Due to its stochastic nature, it often requires the repeated evaluation of expensive computational models, which has in turn created the need for efficient and accurate surrogate modeling strategies.

In this chapter, we provide a review of the state of the art in surrogate modeling for the uncertainty quantification of dynamical systems. 
The specific surrogate modeling choice is connected to the nature of the loading excitation, which can be either fundamentally simple or fundamentally complex (Section~\ref{sec:surrogates:dynamical}). 
Problems with fundamentally simple excitations can often be handled either directly with classical {\it time-frozen} surrogates (Section \ref{sec:surrogates:dynamical}), which approximate each time step independently from the others by introducing principal component analysis for efficiency, or with more advanced techniques such as time or frequency warping (Section \ref{sec:TimeWarping}).
Problems with fundamentally complex inputs require instead different surrogate modeling strategies, based on extensions of non-linear autoregressive modeling (Section~\ref{sec:UQ:NARX}), such as polynomial-chaos NARX (Section~\ref{sec:UQ:PC-NARX}) or the recently developed manifold-NARX (Section~\ref{sec:UQ:mNARX}).

\clearpage

\bibliographystyle{chicago} 
\bibliography{References}

\end{document}